\documentstyle[prl,aps,epsfig,multicol]{revtex}
\begin{document}

\def\tj{$tJ~$}
\def\be{\begin{equation}}
\def\ee{\end{equation}}
\def\bearr{\begin{eqnarray}}
\def\eearr{\end{eqnarray}}
\def\tc{$T_c~$}
\def\tcl{$T_c^{1*}~$}
\def\c2{ CuO$_2~$}
\def\ruo{ RuO$_2~$}
\def\lco{$La_2 Cu O_4~$}
\def\lcso{$La_2 Cu S_2 O_2~$}
\def\lcs{$La_2 Cu S_4 ~$}
\def\bi{bI-2201~}
\def\tl{Tl-2201~}
\def\hg{Hg-1201~}
\def\sro{$Sr_2 Ru O_4$~}
\def\rc{$RuSr_2Gd Cu_2 O_8$~}
\def\mgb{$MgB_2$~}
\def\pz{$p_z$~}
\def\ppi{$p\pi$~}
\def\sqo{$S(q,\omega)$~}
\def\tperp{$t_{\perp}$~}
\def\half{$\frac{1}{2}$~}

\title{Mott Insulator to high \tc Superconductor via Pressure\\
Resonating Valence Bond theory and prediction of new Systems}

\author{ G. Baskaran \\
Institute of Mathematical Sciences\\
C.I.T. Campus,
Madras 600 113, India }

\maketitle

\begin{abstract}
Mott insulator superconductor transition, via pressure and no external
doping, is studied in orbitally non degenerate spin-\half systems. 
It is presented as another RVB route to high \tc superconductivity. 
We propose a `strong coupling' hypothesis which helps to view first 
order Mott transition as a self doping process that also preserves 
superexchange on metal side . We present a generalized t-J model where 
a conserved $N_0$ doubly occupied ($e^-$) sites and $N_0$ empty sites 
($e^+$) hop in the background of $N - 2 N_0$ singly singly occupied 
(neutral) sites in a lattice of N sites. An equivalence to the regular 
t-J model is made and some old and new systems are predicted 
to be candidates for pressure induced high \tc superconductivity.
\end{abstract}

\begin{multicols}{2}[]

\section{Introduction}

Bednorz-Muller's discovery\cite{bednorz} of high temperature 
superconductivity in doped \lco and Anderson's resonating valence 
bond (RVB) theory\cite{pwascience} initiated a new interest in 
Mott insulators as a novel quantum state. In RVB 
theory the pre-existing singlet correlations among electron 
spins in a spin-\half Mott insulator readily become the 
superconducting correlations on doping. The RVB mean field 
theory\cite{pwascience}, 
gauge theory\cite{gbgauge} and later developments\cite{affleck} 
have given results that 
are in qualitative and sometimes quantitative agreement with 
many experimental results.

Motivated by high \tc superconductivity in cuprates, RVB theory has 
so far focussed on the metallization of Mott insulating state by 
external doping. However, we know that there are three families
of `commensurate' tight binding systems that undergo Mott insulator
(spin-Peierls or antiferromagnetic order) to superconductor transition under 
pressure or chemical pressure and no external doping: i) quasi 1 
dimensional $(TMTSF)X_2$, Bechgaard salt family\cite{jerome}
ii) quasi 2 dimensional $\kappa$-$(BEDT$-$TTF)X_2$, 
ET-salt family\cite{osc-book} and 
iii) 3 dimensional fullerites\cite{ramirez,c60-NH3}. For ET and Bechgaard 
salts a single band repulsive Hubbard model at half filling is known 
to be a right model\cite{schulz,fukuyama}.

As antiferromagnetism (more correctly, enhanced singlet 
correlations\cite{hsu})
are present in the insulating side we study Mott transition in spin-\half
orbitally non-degenerate systems from RVB theory point of view.
By looking at a body of experimental results and theories
 on Mott transition\cite{mott} in real systems 
and using the first order character of the Mott transition we 
propose a `{\em strong coupling'} hypothesis; it states that a generic 
Mott transition in real systems is 
to a (strong coupling) metallic state with superexchange. 
This hypothesis allows 
us to view the conducting state as a self doped Mott insulator that 
has very nearly the same superexchange $J$ as the insulator and a 
{\em fixed (conserved) number} $N_0$ of delocalized doubly occupied 
sites and $N_0$ empty sites. This enables us to 
propose a generalized t-J model, where a fixed number $N_0$ of doubly 
occupied sites $(e^-)$  and $N_0$ empty sites $(e^+)$ hop in the
background of $N-2N_0$ singly occupied (neutral) sites that have superexchange
interaction among themselves. Here N is the number of lattice sites.
In determining the total number of mobile charges $2N_0$, that
is the amount of self doping, large range coulomb interaction plays 
an important role.

The issue of RVB superconductivity is solved by transforming our 
generalized t-J model containing $N_0$ holes and $N_0$ doubly occupied sites 
in a Mott insulator into a t-J model that contains either $2N_0$ holes 
or $2N_0$ doubly occupied sites. So our model also exhibits 
superconductivity to the extent the corresponding ordinary t-J model 
exhibits superconductivity. Encouraged by our theory we make certain 
predictions about possibility of pressure induced superconductivity 
in a family of compounds: i) old ones such as three dimensional $CuO$, 
layered \lco, infinite layer $CaCuO_4$, insulating Tl and Hg cuprates 
and YBCO and ii) new ones such as \lcso, \lcs, $CaCuS_2$ with $CuS_2$ 
planes or their selenium analogues, to mimic chemical
pressure along the ab-plane. 

It should be pointed out that, 1d Mott transition and various Hubbard 
model based theories exist in the 
literature\cite{schulz,fukuyama,tosatti} for the Bechgaard, ET salts
and fullerites. Our view point emerging from {\em `strong coupling'} 
hypothesis and the resulting generalized t-J model emphasizes 
that the physics of the conducting state is also determined by a 
strong coupling physics with superexchange  and the consequent RVB
physics.  

Standard thought experiment of Mott transition is an adiabatic 
expansion of a cubic lattice of hydrogen atoms forming a metal. 
Electron density decreases on expansion and Thomas-Fermi screening 
length increases; when it becomes large enough to 
form the first electron-hole bound state, there is a first order 
transition to a Mott insulating state, at a critical value of the 
lattice parameter $a \approx  4a_B$, where $a_B$ is the Bohr radius.
The charge gap jumps up from zero to a finite Mott-Hubbard gap
across the transition (figure 1a), by a feedback process that 
critically depends on the long range part of the coulomb interaction, 
as emphasized by Mott\cite{mott}. 

Experimentally known Mott transitions are first order and the 
insulating side close to the transition point usually have a substantial 
Mott-Hubbard gap; in oxides this gap is often of the order of an eV. 
In organics, where 
the band width are narrow $\approx 0.25 eV$ the Mott Hubbard gap 
also has similar value. In view of the finite Mott Hubbard gap, the 
magnetism on the Mott insulating side is well described by an effective
Heisenberg model with short range superexchange interactions. There are
no low energy charge carrying excitations. That is, we have a strong 
coupling situation.

\begin{figure}[h]
\epsfxsize 8cm
\centerline {\epsfbox{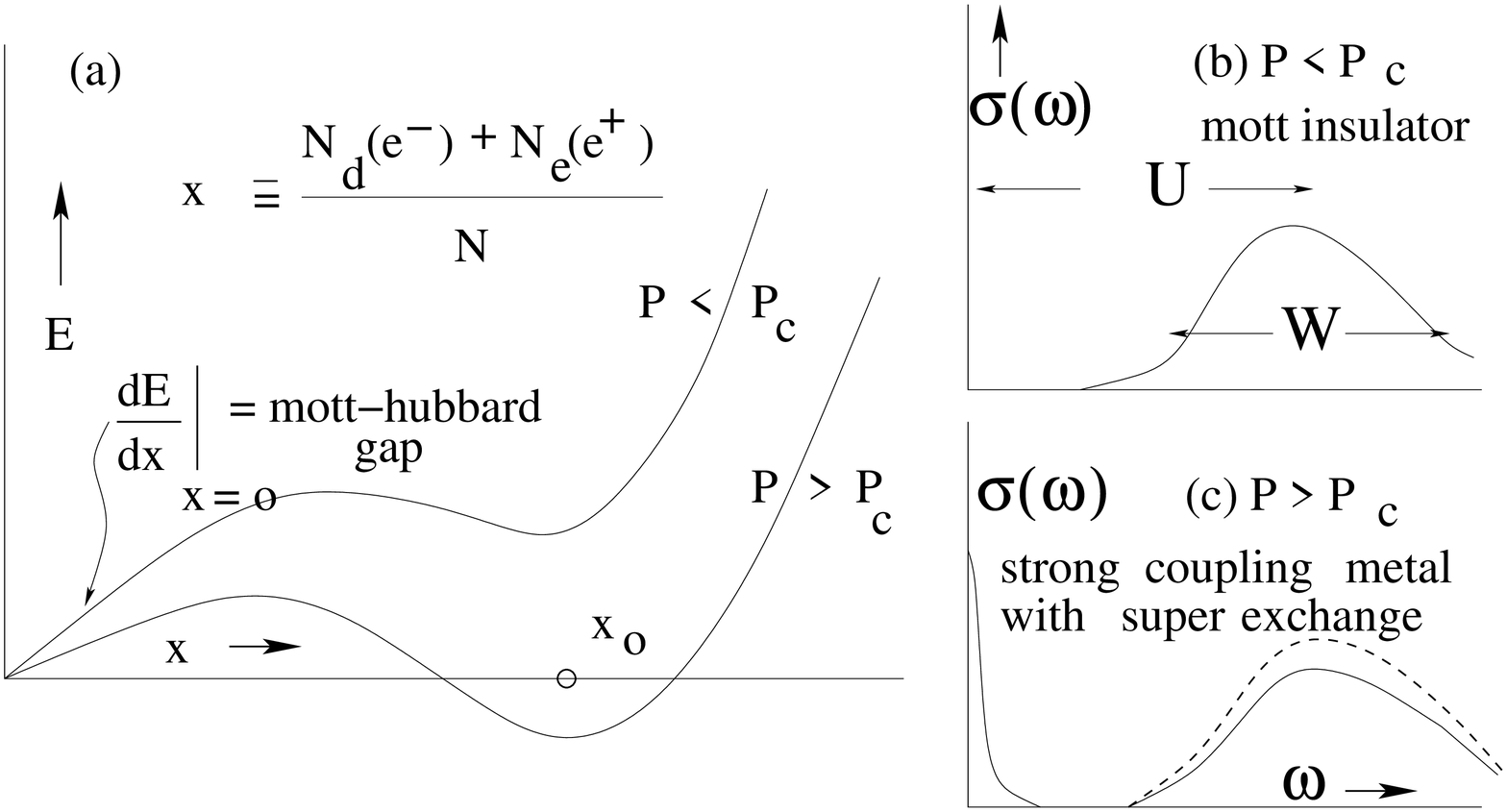}}
\caption{ a) Energy of a half filled band above and below the
critical pressure $P_c$, as a function of 
$x = \frac{N_d(e^-) + N_e(e^+)}{N}$. Here $N_d(e^-) = N_e(e^+)$ are 
the number of doubly occupied ($e^-$) and number of empty sites ($e^+$); 
total number of lattice sites N = total number of electrons. 
Optimal carrier density $x_0 \equiv \frac{2N_0}{N}$ is determined by long 
range part of coulomb interaction and superexchange energy. 
b) and c) Schematic picture of the real part of the frequency 
dependent conductivity on the insulating and metallic side close
to the Mott transition point in a real system. W is the band width.}
\end{figure}

What is interesting is that this strong
coupling situation continues on the metallic side as shown by optical 
conductivity studies for example in Bechgaard\cite{digiorgi} and ET salts: 
one sees 
a very clear broad peak (a high energy feature) corresponding to the 
{\em upper Hubbard band both in the insulating and conducting states}. 
The only difference in the conducting state is the appearance of Drude 
peak, whose strength and shape gives an idea of number of free carriers 
that have been liberated (figure 1b and 1c). As the location and width of the 
Hubbard band has only a small change across the transition, one may 
conclude that the local quantum chemical parameters such as the hopping 
matrix elements t's and Hubbard U (corresponding superexchange J) remain 
roughly the same.  
This is the basis of our `strong coupling' hypothesis:  
{\em a generic Mott insulator metal transition in real system is to a 
(strong coupling) metallic state that contains superexchange}. 

As superexchange survives in the conducting state, two neighboring singly 
occupied sites of net charge $(0,0)$ can not decay into 
freely moving doubly occupied and empty sites $(e^-,e^+)$. 
Conversely a pair of freely moving doubly occupied and empty 
sites cannot annihilate each other and produce a bond singlet (figure 2).
(Recall that in a free fermi gas, where there is no superexchange, 
the above processes freely occur). Superexchange and long range part
of the coulomb interactions determine the number of self doped carriers
$2N_0$ and their conservation. 
\begin{figure}[h]
\epsfxsize 6cm
\centerline {\epsfbox{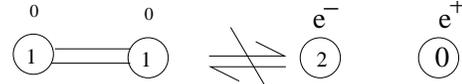}}
\caption{If superexchange survives on the metallic side,
a pair of neighboring singly occupied sites can not decay 
into freely moving doubly occupied and empty sites. The converse 
is also true.}
\end{figure}
The above arguments naturally leads to 
a generalized t-J model for the conducting side in the vicinity of 
the Mott transition point 
\bearr
H_{tJ}  =  & - & \sum_{ij}t^{}_{ij}  P_d~c^\dagger_{i\sigma} 
c^{}_{j\sigma} P_d
     -  \sum_{ij}t^{}_{ij}  P_e~ c^\dagger_{i\sigma} c^{}_{j\sigma} P_e
 +  h.c.  \nonumber \\
      & - & \sum_{ij} J_{ij} ( {\bf S}_i \cdot {\bf S}_j - 
\frac{1}{4} n^{}_i n^{}_j ), 
\eearr
operating in a subspace that contains a fixed number $N_0$ of 
doubly occupied and $N_0$ empty sites. The projection
operators $P_d$ and $P_e$ allows for the hopping of a doubly occupied 
and empty sites respectively in the background $N - 2N_0$ of 
singly occupied sites. Here $N$ is the total number of electrons, 
which is the same as the number of lattice sites. As the Mott-Hubbard
gap is the smallest at the Mott transition point, higher order 
superexchange processes may also become important and contribute to
substantial non neighbor $J_{ij}$'s.

Our t-J model adapted to the self doped Mott insulator has a more
transparent form in the slave boson representation $ c^{\dagger}_{i\sigma}
\equiv s^{\dagger}_{i\sigma}d^{}_{i} + \sigma s^{}_{i{\bar \sigma}}
e^{\dagger}_{i} $. Here the chargeons
$d^{\dagger}_i$'s and $e^{\dagger}_i$'s are hard core bosons that create
doubly occupied sites ($e^-$)  and empty sites ($e^+$) respectively. 
The fermionic
spinon operators $s^{\dagger}_{i\sigma} $'s create singly occupied sites
with a spin projection $\sigma$. The local
constraint, $ d^{\dagger}_{i} d^{}_{i} + e^{\dagger}_{i} e^{}_{i} +
\sum_{\sigma} s^{\dagger}_{i\sigma} s^{}_{\sigma} = 1 $, keeps us in 
the right Hilbert space.

In the slave boson representation our t-J model takes a suggestive form:
\bearr
 H_{tJ}  & = &  - \sum_{ij}t^{}_{ij}
( d^{\dagger}_{i} d^{}_{j} 
\sum_\sigma s^{}_{i\sigma} s^{\dagger}_{j\sigma} 
+
e^{}_{i} e^{\dagger}_{j} \sum_\sigma 
s^{\dagger}_{i\sigma} s^{}_{j\sigma} ) + h.c.
\nonumber \\
& - & \sum_{ij} J_{ij} b^{\dagger}_{ij}b^{~}_{ij}
\eearr
where $b^{\dagger}_{ij} = \frac{1}{\sqrt 2} ( s^\dagger_{i\uparrow} 
s^\dagger_{j\downarrow} -  s^\dagger_{i\downarrow} 
s^\dagger_{j\uparrow})$ is a spin singlet spinon pair creation 
operator at the bond $ij$. It is easily seen that the total
number operator for doubly occupied sites $\hat{N}_d \equiv \sum 
d^{\dagger}_{i} d^{~}_{i}$ and empty sites $\hat{N}_e \equiv \sum 
e^{\dagger}_{i} e^{~}_{i}$ commute with the t-J Hamiltonian 
(equation 2):
\be 
\left[ H_{tJ}, \hat{N}_d \right] =  
\left[ H_{tJ}, \hat{N}_e \right] = 0
\ee  
That is, $\hat{N}_d$ and $\hat{N}_e$ are individually conserved. In our 
half filled band case $N_d =N_e = N_0$. (This special conservation law
is true only for our effective t-J Hamiltonian and not for the original
Hubbard model). 

This conservation law allows us to make 
the following statement, which is {\em exact for a particle-hole symmetric
Hamiltonian and approximate for the asymmetric case}: our generalized
t-J model with 
a fixed number $N_0$ of doubly occupied sites and equal number $N_0$ of 
empty sites has the same many body spectrum as the regular t-J model 
that contains either $2N_0$  holes or $2N_0$electrons. Symbolically it means 
that $H_{tJ}(N_0,N_0) \equiv H_{tJ}(2N_0,0) \equiv H_{tJ}(0,2N_0)$. 
This means we can borrow all the known results of t-J model, viz. 
mean field theory, variational approach, numerical approach
etc. and apply to understand the thermodynamic and superconductivity 
properties of our self doped Mott insulator. Response to electric and
magnetic field perturbation has to be done separately as the $d$ 
and $e$ bosons carry different charges, $e^-$ and $e^+$ respectively.

Another consequence of the above equivalence is shown in figure 3,
where we have managed to draw the path of pressure-induced Mott 
transition in a Hubbard model phase diagram, even though Hubbard model
does not contain the crucial long range interaction physics. The jump 
from B to C is the first order phase transition, remembering that in 
the presence of our new conservation law what decides the spectrum 
of our generalized t-J model is the total number of $e^+$ and $e^-$ 
charge carriers in an equivalent regular  t-J model. 
The horizontal jump is also 
consistent with our strong coupling hypothesis.

\begin{figure}[h]
\epsfxsize 7cm
\centerline {\epsfbox{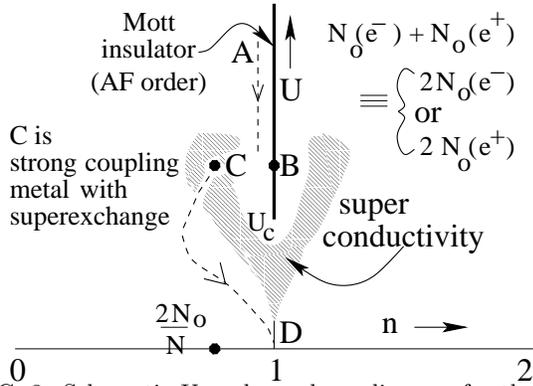}}
\caption{Schematic U-n plane phase diagram for the Hubbard model.
ABCD represents the path a real system takes as pressure
increases. B to C is the first order Mott transition, consistent
with our {\em strong coupling} hypothesis. The point C, from a 
regular t-J model point of view, is hole doped at density 
$n = \frac{2N_0}{N}$; however, based on our equivalence it corresponds 
to a half filled band with a total of $N_0(e^-) $ + $N_0(e^+)$ self
doped carriers.}
\end{figure}

An important parameter in our modeling is the equilibrium total 
$ e^+$ and $e^-$ carrier concentration, $ x_0 \equiv \frac{2N_0}{N}$ 
in our self doped Mott insulator. This also controls the value of
superconducting \tc we will get across the Mott transition point.
Estimate of $x_0$ depends on the long range part of the coulomb
interaction energy and also the short range superexchange energy;
we will defer this discussion to a later publication. $x_0$ 
may also be determined from experiments such as frequency dependent 
conductivity by a Drude peak analysis. 

Since we have reduced our self 
doped Mott insulator problem into a t-J model, superconducting \tc 
is determined by $t, J ~{\rm and~} x_0$, as in the t-J model. If 
exchange interaction contribution is comparable to the long range
coulomb contribution, $x_0$ will be closer to value that maximizes
superconducting \tc. Another important point is the possibility
of non nearest neighbor superexchange $J_{ij}$ processes, 
which i) frustrate long range antiferromagnetic order to encourage 
spin liquid phase and ii) increase 
the superexchange energy contribution to the total energy; this could 
give a larger superconducting \tc across the Mott transition than 
expected from a t-J model with nearest neighbor superexchange. 
Perhaps an optimal self doping and sufficiently frustrated superexchange 
interactions is realized in $(NH_3)K_3C_{60}$ family\cite{c60-NH3}, 
since Neel temperature 
$T_n \approx 40 K$ and superconducting \tc $\approx 30 K$ are 
comparable. 

If the self doping is small there will be competition from 
antiferromagnetic metallic phase, stripes and phase separation.
For a range of doping one may also get superconductivity from 
inter plane/chain charge disproportionation. If self doping is very 
large then the effect 
of superexchange physics and the consequent local singlet correlations 
are diluted and the superconducting \tc will become low. This is the 
reason for the fast decrease of superconducting \tc with pressure 
in the organics. 

In what follows we discuss some families of compounds, some old
ones and some new ones and predict
them to be potential high \tc superconductors, {\em unless some 
crystallographic transitions or band crossing intervenes} and change 
the valence electron physics drastically. 
$CuO$, is the mother compound\cite{cuo}
of the cuprate high \tc family. It is 
monoclinic and $CuO_2$ ribbons form a 3 dimensional network, 
each oxygen being shared by two ribbons mutually perpendicular to each other. 
The  square planar character from four oxygens surrounding a Cu in a 
ribbon isolates one non-degenerate valence $d$-orbital with a lone electron. 
This makes $CuO$ an orbitally non-degenerate spin-\half Mott insulator 
and makes it a potential candidate for our pressure route to 
high \tc superconductivity. The frustrated superexchange
leads to a complex three dimensional magnetic order with a Neel 
temperature $\sim 230 K$. These frustrations should help in 
stabilizing short range singlet correlations, which will help
in singlet cooper pair delocalization on metallization.  

As far as electronic structure is concerned, the $CuO_2$ ribbons
give $CuO$ a character of coupled 1d chains. This makes it some what 
similar to quasi one dimensional Bechgaard salts, which has a Mott 
insulator to superconductor transition, via an intermediate metallic
antiferromagnetic state as a function of physical or chemical pressure. 
The intermediate metallic antiferromagnetic state represents a 
successful competition from nesting instabilities of flat fermi 
surfaces arising from the quasi one dimensional character. Once 
the quasi one dimensional character is reduced by pressure, nesting 
of fermi surface is also reduced and the RVB superconductivity takes
over.

If manganite\cite{mang}, a perovskite and fullerites\cite{c60-NH3}
are any guidance, metallization 
should take place under a pressure of $\sim$  tens of GPa's. $CuO$ should 
undergo a Mott insulator superconductor transition, perhaps with an 
intermediate antiferromagnetic metallic state. The superconducting
\tc will be a finite fraction of the Neel temperature, as is the
case with Bechgaard salts or $K_3 (NH_3) C_{60}$. Thus an optimistic
estimate of \tc will be 50 to 100 K.

Similar statements can be made of the more familiar \lco, insulating
YBCO and the $CaCuO_2$, the infinite layer compound or the family 
of Mott insulating cuprates such as Hg and Tl based insulating 
cuprates. Infinite layer compound has the advantage
of absence of apical oxygen and should be less prone to serious
structural modifications in the pressure range of interest to us.
The quasi 2d Hubbard model describing the $CuO_2$ planes does have 
an appreciable $t'$, making nesting magnetic instabilities weaker. 
Thus we expect that on metallization a superconducting state to be 
stabilized with a small or no antiferromagnetic metallic intermediate 
state. 

The quasi 2d cuprates have a special advantage in the sense we may
selectively apply ab-plane pressure in thin films by epitaxial
mismatch and ab plane compression. Apart from regular pressure   
methods, this method\cite{locquet} should be also tried. 

One way of applying chemical pressure in cuprates is to increase
the effective electron band width by increasing the band parameters
such as $t$ and $t'$ in the Hubbard model. This can be achieved by
replacing oxygens in the $CuO_2$ planes (or in 3 dimensional $CuO$) 
by either sulfur or selenium,
which because of the larger size of the bridging 3p or 4p orbitals
increase the band width and at the same time should reduce the 
charge transfer or Mott-Hubbard gap. On partial replacement of oxygen, 
as $CuO_{2-x}X_x$ in the planes or $CuO_{1-x}X_x$ ($X = S, Se$)
one might achieve metalization without doping. 

Some possible new stoichiometric compounds are \lcso, \lcs 
and $CaCuS_2$ or their Se
versions. Synthesizing these compounds may not be simple, as
the filled and deep bonding state of oxygen $2p$ orbitals in $CuO_2$ 
play a vital role in stabilizing square planar coordination. 
With S or Se versions these bands will float up and come closer 
to the fermi level thereby making square structure less stable. 
Under pressure or some other non equilibrium conditions some 
metastable versions of these compounds may be produced. 
One could also optimize superconducting \tc by a judicious 
combination of pressure induced self doping and external doping.

I thank Erio Tosatti for bringing to my attention pressure induced
Mott insulator superconductor transition in $(NH_3)K_3C_{60}$ and for 
discussions.

\end{multicols}
\end{document}